# Perturbation waves in proteins and protein networks:
## Applications of percolation and game theories in signaling and drug design

Miklós A. Antal, Csaba Böde and Peter Csermely*

*Department of Medical Chemistry, Semmelweis University, H-1088 Budapest, Puskin str. 9, Hungary*

**Abstract:** The network paradigm is increasingly used to describe the dynamics of complex systems. Here we review the current results and propose future development areas in the assessment of perturbation waves, i.e. propagating structural changes in amino acid networks building individual protein molecules and in protein-protein interaction networks (interactomes). We assess the possibilities and critically review the initial attempts for the application of game theory to the often rather complicated process, when two protein molecules approach each other, mutually adjust their conformations via multiple communication steps and finally, bind to each other. We also summarize available data on the application of percolation theory for the prediction of amino acid network- and interactome-dynamics. Furthermore, we give an overview of the dissection of signals and noise in the cellular context of various perturbations. Finally, we propose possible applications of the reviewed methodologies in drug design.

Key words: allosteric regulation; conformational changes; drug targets; graphs; hubs; network dynamics; protein folding; signaling

## 1. INTRODUCTION: AMINO ACID NETWORKS AND INTERACTOMES

The network concept gained an increasing ground in the analysis and prediction of complex system behavior in the last decade. Networks help our insight and understanding by reducing the complex system to a set of interacting elements{nodes/vertices}[†], which are bound together by links{contacts/edges/interactions}. In most currently used networks{graphs} links represent interactions between element pairs. Links usually have a weight, which characterizes their strength{affinity/intensity/probability}. Links may also be directed, when one of the elements has a larger influence to the other than vice versa [1-3].

The complexity of the living cell can be approached by several structural and functional networks. Cellular functions are well-described by metabolic networks, where the elements are small metabolites, and the links are the enzymes which produce them. In signaling and gene regulatory networks the signaling molecules constitute the elements, while their interactions give the links. Amino acid networks and protein-protein interaction networks{interactomes} form the two basal layers of the hierarchical networks of cellular structure. In the most commonly used amino acid networks the elements are the amino acids of protein molecules, while the links represent their neighboring position in space, if the inter-element distance is below a cut-off (which is usually between 0.45 and 0.85 nm). Amino acid networks may use weights instead of the cut-off, and may also discriminate individual atoms of the protein structure as elements [1, 3, 4].

Protein-protein interaction networks catalogue the interaction of cellular proteins. Regretfully there are only a few initial attempts to provide detailed information of the interaction strength [5] as well as the variations present in individual, single cells [6]. Currently, most protein interactomes contain a list of the most probable protein-protein interactions in an average cell of the respective organism, where link weights (if exist) represent the probability of the particular interaction. This interaction-probability roughly correlates to the association constant of the given protein pair [7] but a detailed characterization of interactomes is clearly an important task of future research. Higher levels of the structural hierarchy in the cell (going beyond the scope of our current review) can be described by the cytoskeletal and organelle-membrane networks [3].

Most cellular networks are small worlds, where two elements of the network are separated by only a few other elements. Small-worldness helps the fast transmission of perturbations. Networks contain hubs, i.e. elements, which have a large number of neighbors{have a high degree}. Amino acid networks have a Poissonian degree distribution, which means that they have a negligible amount of hubs. On the contrary, protein-protein interaction networks often display scale-free degree distribution, which means that the probability to find a hub with a number of neighbors a magnitude higher is a magnitude lower (but, importantly, not negligible). Both amino acid and protein-protein interaction networks can be dissected to overlapping modules{communities/groups}, which often form a hierarchical structure. Both hubs and modules provide a filtering mechanism to prevent perturbation-overload and to avoid the excessive propagation of network damage. Amino acid networks and interactomes are often heterogeneous, and their different modules may behave completely differently. Moreover, sampling bias and improper data analysis may show small-worldness, scale-free distributions and modularity in such cases, where they do not actually exist. Therefore, special caution has to be taken to scrutinize the validity and extent of datasets, use correct sampling procedures and adequate methods of data analysis [1,3,4,8].

---

*Address all correspondence to this author at the Department of Medical Chemistry, Semmelweis University, H-1088 Budapest, Puskin str. 9, Hungary; Tel: +36-1-266-2755; Fax: +36-1-266-6550; E-mail: csermely@puskin.sote.hu
[†]In a few cases we list a number of commonly used synonyms of the given word to make it more familiar to those coming from various sub-disciplines.

## 2. DYNAMICS OF AMINO ACID NETWORKS

Protein structure dynamics plays an essential role in protein function and regulation. The role of dynamics has been studied both theoretically [9] and experimentally [10,11] since a long time. Several, methodologically different approaches exist, which use network methodology, like (i) the energy network models, (ii) models using mostly physical and statistical mechanical approaches like Gaussian network models (GNM) [9] and (iii) network methods using information and graph theory approaches, like the protein structural networks [12].

Amino acid networks take into account only the interactions between amino acid side-chains, and neglect the constraints of the protein backbone. The neglect of the protein backbone does not make a problem, if we analyze only the topology of these networks, and want to draw conclusions for the structure and stability of proteins. As an example for this, using a structural network, Atilgan et al. [12] showed that the fluctuations of amino acid side chains (taken from experimental data) are strongly correlated with the spatial arrangement of protein residues. This reflects that central amino acids (having a smaller average of their shortest path lengths) have a more restricted motion.

Different methods were used to understand the dynamics of topological networks and to explain protein motions and conformational rearrangements. One possibility is the elastic network model, where only the atomic coordinates of the αC atoms are used to build the network, which makes the calculations computationally inexpensive. In this model a harmonic potential is used to account for pair-wise interactions between all αC atoms [9,13]. However, such a network cannot be studied by the mathematical framework of graph theory, and it requires more sophisticated statistical mechanical methods. Using the elastic network model a set of sparsely connected, highly conserved residues were identified, which are key elements for the transmission of allosteric signals in three nanomachines, such as DNA polymerase, myosin and the GroEL chaperonin [13]. Protein backbone motions had been predicted for a set of proteins and showed good agreement with experimental results, when the reorientational contact-weighted elastic network model was applied [14].

Another elastic network representation includes all atoms, forming a spring network [15]. Overconstrained (having more crosslinking bonds than needed) and underconstrained (with less crosslinking bonds than needed) protein regions were successfully identified using this approach. These regions correspond with rigid and flexible protein segments, respectively [14]. A recent paper combines the elastic network model with a network-theory approach, underlying the observation that functionally active residues have enhanced communication (connection) properties [16]. Since this model investigates the information propagation time, this approach may shed new insight on the allosteric function of enzymes.

Networks can also be used to model conformational transitions. In energy network models nodes represent conformational states of the protein, while links correspond to the transitional states between them [17]. An interesting Monte Carlo study by Andrec et al. [18] combines the results of molecular dynamics simulation with network approach to understand the folding kinetics of the G-protein C-terminal peptide. In this model the conformational states were approximated by replica-exchange molecular dynamics simulations, and the transitions were studied by network methodology. With this method helical on-pathway intermediates had been observed during folding the G-peptide β-hairpin. In similar studies the folding kinetics of the villin headpiece has also been investigated. By constructing the Markovian State Model it became possible to propagate villin dynamics to times far beyond the directly simulated, and to rapidly calculate long time kinetics (to tens of microseconds) and evolution of ensemble property distributions [19]. Although due to computational restrictions only small proteins (peptides) were studied so far using the network approach, this area will certainly provide a number of interesting results in the near future.

The energy network approach is analogous to that of the conformational networks. In this latter approach the energy landscape of the protein is modeled by a network: the nodes of are the different conformational states of the protein, while the links correspond to the transition states between them. The energy landscape has both a small-world and scale-free character [20,21], which might give an explanation of the high dynamism of most protein structures: the small-world character ensures that a node of network (which always represents a protein conformation) is only a few steps (conformational transitions) apart from any other conformation. Besides explaining the large flexibility in protein function and regulation, the small-world character, when applied to protein-folding, also provides an alternative, network-based explanation to solve the Levinthal-paradox [4].

**3. PROTEIN-PROTEIN INTERACTIONS: A POSSIBLE APPLICATION OF GAME THEORY**
When a protein-protein interaction develops, the two partners approaching each other continuously interact with each other influencing each other's structure (or in another framework each other's energy landscape). Several models had been proposed to describe the process of protein-protein interaction, beginning with the well known "lock and key model" proposed by Emil Fischer in the $19^{th}$ century [22]. However, it is clear, that most protein-protein interactions cannot be described in such a simple, rigid and one-step way. Several pieces of evidence suggest that the rigidity of the "lock and key model" is not a good approximation, since during the interaction process proteins influence each other's structure. The "induced fit" model [23] describing this conformational interdependence had been successful for a bundle of proteins [24] and is still a centerpiece of our biochemical understanding of protein interactions and enzyme function.

Recently an alternative for the induced fit mechanism, the "pre-existing equilibrium/conformational selection" model emerged. For this model it had been proposed that the native state of the protein cannot be described as one well-defined conformational state, but rather reflects a conformational ensemble, from which the most suitable conformational state(s) binds the other protein (or the substrate) therefore shifting the equilibrium towards the complex formation. Suitable, 'binding-competent' conformational states are often well-populated, likely conformations of the original, 'lonely' protein structure, which helps a lot to 'lock-in' the protein to the 'binding-competent' conformation [25,26].

However, from both the experimental evidence and from theoretical results it seems that the protein-protein interaction process is much more complicated than the above models suggest. When two proteins bind to each other, neither of them can be approximated as a small, rather rigid molecule, which makes this scenario much more complex, than the "simplified" substrate-binding models (in fact the substrate molecules are also flexible, therefore, in principle the ideas below also hold on the explanation of the molecular details of enzyme kinetics). Here we list a few recent observations, which suggest that the development of protein-protein interactions is a multi-step, sequential process, where many consecutive steps can only happen, if certain preceding steps have been successful, and where 'binding-competent' steps from one of the binding proteins require cooperative, preceding or successive 'binding-competent' steps from the other binding protein. This 'interdependent protein dance' can be described well in terms of the game theory [27].
- In many cases one-step models do not explain well the sequential and multi-dependent conformational changes which take place during protein-complex formation [28].
- At the interaction of cytochrome-c with lysosime a weak, long-range attraction had been observed, which had a range several times that of the diameter of the protein molecule. This interaction enables the development of 'game-steps' and allows the development of an intricate communication pattern, an 'approach-path' as the two proteins become increasingly engaged in the interaction. Moreover, the interaction was strongly influenced by the ions present in the solution, which introduced yet another set of players into the already complex protein-protein game [29].
- During the initial steps of protein-protein interactions a number of transient complexes are formed. However, in most cases such transient states escaped detection by usually applied experimental techniques. In recent studies the assembly of the 30S ribosomal subunit was assessed in detail and it has been shown that the protein-RNA complex undergoes various local conformational transitions as the assembly develops [30]. In another study using paramagnetic relaxation enhancement an ensemble of various transient, non-specific encounter complexes was observed during the encounter of several protein complexes including the amino-terminal domain of enzyme I and the phosphor-carrier protein HPr [31].

The game-theory approach ('protein-games') [27] may give a novel insight to understand the complex phenomenon of protein-protein interactions. The emergence of cooperation is a long-time studied field in game-theory [32]. Conditions helping the cooperation in spatial games [33] offer a very helpful framework to apply the results of game theoretical studies to the formation of protein-protein complexes. Since spherical constraints significantly reduce the possible network topologies of the two amino acid networks (i.e. proteins) 'playing with each other', general conditions extending cooperation can be helpful to predict key requirements of successful protein-protein interactions from the point of the interdependent protein dynamics. A recent report highlighted two basic conditions, learning and innovation, to extend the network topologies able to maintain a significant level of cooperation [34]. In protein-protein interactions, learning may correspond to the steering process as the two proteins gradually approach each other, and gain an increasing amount of information of the other's structure and requirements for efficient docking, like in the case of cytochrome c and lysosime we mentioned above [29]. Innovation is actually a low level of randomness, which is emerging from all the protein dynamics we described above. In a recent summary starting from the significant presence of structural disorder in protein complexes, Tompa and Fuxreiter [35] raised the general possibility of "fuzzy complexes", where a non-significant disorder is a general feature of the protein complex. Such fuzzy conditions may also significantly contribute to the level of innovation, which is needed to maintain cooperation at a wider range of network topologies [34], which is a usual requirement during protein complex formation, where both amino acid networks engaged in this process undergo a set of significant topological changes.

**4. INTERACTOME-DYNAMICS**
Protein-protein interaction networks display a high dynamism. The seemingly 'rock-solid' core-histones, protected by both the rest of the nucleosomal structure and by the DNA wrapped around them, have a surprisingly little 5 minutes half-life only in their original position. The cell is full with 'moonlighting' proteins, which appear at completely unexpected positions and functions [36-38]. Beyond the continuous link re-arrangements, proteolysis and synthesis of cellular proteins make to vanish and re-appear a lot of interactome elements. Network modules of the yeast interactome may be dissected to static and dynamic modules using the information of gene expression changes. The pathway structure of static modules is more redundant, which allows a faster evolution and larger tolerance of gene expression noise. On the contrary, dynamic modules help the condition-dependent, flexible regulation of cellular responses [39,40]. Different forms of protein dynamics can be easily discriminated in case of date-hubs and party-hubs, where date-hubs form complexes with different subsets of their partners at different times and cellular locations, while party-hubs form complexes with all of their partners simultaneously. Date hubs – logically – usually have a single binding surface, while party hubs are multi-interface proteins. Date hubs contain more disordered regions, while party hubs have a larger tendency to form a rich-club, i.e. a network region, where party-hubs are associated preferentially with each other [40-43].

**5. APPLICATIONS OF PERCOLATION THEORY TO ASSESS NETWORK TOPOLOGY, DYNAMICS AND EVOLUTION**
Percolation theory is a widely used model in rather different areas from porous material characteristics, as well as transport theory to the spread of information or modeling the propagation of forest fires [44,45]. The renaissance of network research triggered several percolation-related studies in a number of new directions including amino-acid networks and protein interactomes, where the percolation phenomenon, i.e. the emergence of a large, communicating network segment can efficiently model a number of key biological processes including protein folding, network rearrangements in stress and disease, and the transmission of information within and between proteins. Additionally, the percolation-based assessment of system-level parameters (emergent properties) of cellular networks helps to understand cell signaling, differentiation, death and evolution. However, we must warn that we are in a very early stage of connecting protein networks and percolation theory. In the following part we will summarize the initial opportunities and highlight further areas of exciting studies.

The application of percolation theory is especially useful in situations, where the formation of amino-acid networks can be observed, i.e. during the protein folding process [46]. Self-similarity and fractal-like structure is typical to many real-world networks [47]. The amino acid network of protein structure is also self-similar to a certain level, and the resulting protein structure can be characterized by a fractal dimension, a quantity describing the roughness of the surface [48]. In protein folding percolation is achieved, when a giant-component of the amino-acid network is formed, i.e. when at a critical point most of the amino acids become abruptly connected as folding proceeds. Formation or destruction of the giant component can be characterized by a dynamic scaling behavior, i.e. the approximation of relevant physical properties, like correlation length and free energy, by power-laws in the vicinity of the critical point. In this approximation the difference of the physical property from the critical point values is raised to the power of the critical exponents. These critical exponents are thus crucial, as they characterize well the actual values of many key functions of protein folding. Some of the critical exponents describing the folding process were found to depend only on the fractal dimension and on the Euclidean dimensionality [49]. This direct relation between the static network properties and the dynamic network behavior is an important contribution to both percolation theory and network science. However, proteins display a multi-fractal behavior: it is difficult to define a single fractal dimension for a large protein molecule, because of its non-homogeneous structure and of the absence of complete self-similarity [50]. Nevertheless, protein folding offers the possibility for a number of percolation-based further studies. Protein function and the interactions of folding proteins (e.g. with surface water) may undergo abrupt changes when the underlying amino acid network gets close to the critical point. Percolation-studies on local network segments (e.g. modules of the amino acid network often corresponding to functional protein modules or domains [4]) may offer an additional level of understanding, where the multi-fractality of proteins may be better approximated by single, local fractal dimensions.

Besides the possible modularity of amino acids networks, studies monitoring the breakdown of percolation may also be used to identify modules of protein-protein interaction networks. For this purpose, interaction patterns are represented in the adjacency matrix, where the entry is $A_{ij} = 1$, if the nodes $i$ and $j$ are linked, while otherwise $A_{ij} = 0$. The difference between eigenvalues of the adjacency matrix is called level spacing. Depending on correlations between eigenvalues, the level spacing distribution can follow the statistics of Gaussian orthogonal ensemble (strong correlation), Brody distribution (intermediate state) or Poisson distribution (no correlation). It has been shown for random networks that level spacing distribution changes from Gaussian statistics through Brody distribution to Poisson distribution, as the critical point is approached by the decrease of the average degree [51]. This observation holds for the protein-protein interaction network of *Saccharomyces cerevisiae*: when the giant component is destroyed in a sequential node deletion process simulating an intentional attack, the level spacing distribution undergoes a dramatic transition. In this process the Gaussian statistics of the level spacing distribution is replaced by exponential curves referring to Poissonian statistics. As percolation is accompanied by the above changes in eigenvalue distributions, it is possible to identify modules of the protein interactome containing elements, which remain linked even at the level of link-removal, when the Poissonian statistics appear [52].

Beyond the identification of protein-protein interaction modules by percolation studies, modular differences in the percolation process offer an efficient method for the characterization of real-world protein networks and may also reflect important functional consequences. Different local levels of percolation may reveal different functional states of cellular modules/protein complexes, which may undergo profound changes in stress, during the cell cycle, cell differentiation and disease.

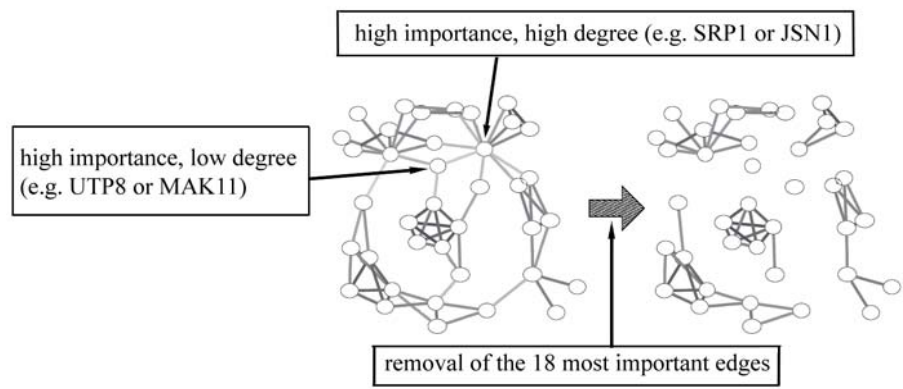

**Fig. (1).** Percolation analysis of the yeast protein-protein interaction network. The illustrative figure summarizes some of the finding of Chin and Samanta [49]. Changes in percolation identified two classes of important proteins. The first class consisted of hubs having a high degree, like the nuclear localization signal receptor protein, SRP1 or the mitochondrial receptor, JSN1. Proteins of the second class had high betweenness but low average degree, like the tRNA nuclear exporter, UTP8 or the pre-ribosomal factor, MAK11 [49].

Another paper with a strong mathematical background, but with more direct applications uses bond-percolation to define a novel measure of significance of proteins in their cellular context. Originally, the importance of proteins measured by their essentiality was correlated by their number of neighbors, i.e. degree [53]. The percolation-based study of Chin and Samanta [49] used global connectivity measures in the unweighted yeast protein-protein interaction network, which were shown to correlate stronger with essentiality than the local connectivity data (degrees) of the individual proteins. To define global connectivity a stochastic analysis was made by randomly removing a given proportion of edges. The importance of a vertex was given by the fraction of other vertices to which it remained linked; importance of links was measured by their contribution to the overall connectedness (i.e. the number of all protein pairs that were connected in their presence but became disjoint without them) (Fig. **1**). These definitions seem to be rather useful for characterizing biological significance and agree well with the results of other studies correlating the importance with betweenness centrality [54] and a centrality measure coming from the overlapping modular structure of the interactome [55]. Percolation-based significance measures may be conducive for important studies in other biological systems, such as in the determination of hot-spots in amino-acid networks, or those of combined amino-acid/water structural networks of proteins or protein complexes.

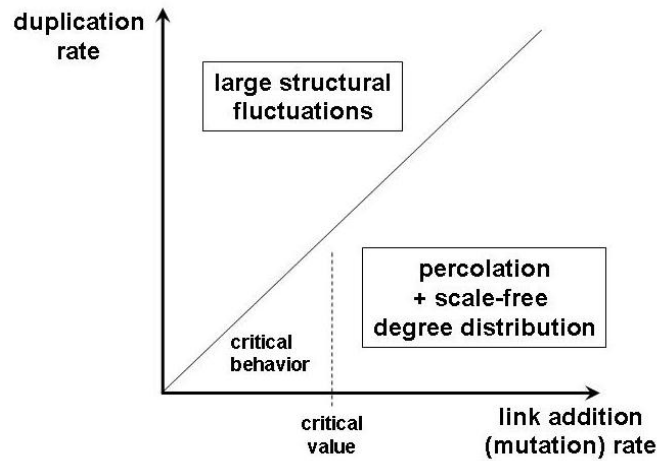

**Fig. (2).** A schematic 'phase-diagram' of a percolation-based protein interaction network evolution model. The model of Kim et al. [56] used the dual effects of the functional duplication of proteins and the addition of random links representing possible mutations. When link-addition was dominant, an infinite-order percolation transition arose at a critical value of the addition. In the opposite extreme of the parameter-set, when the duplication rate was extremely high, the network exhibited giant structural fluctuations in different realizations. The study showed that mutations are vital for self-averaging and the emergence of robustness and statistical properties similar to those observed in real protein interaction networks.

A third study connecting percolation theory and protein-protein interactomes considers an evolving protein interaction network with functional duplication of proteins and the addition of random links representing the possible mutations [56]. When link-addition was dominant, an infinite-order percolation transition arose at a critical value of the addition rate (Fig. **2**). The link addition rate appears in the differential equation from which the expected number of percolating clusters can be obtained. Thus, link addition affects both the expected number of percolating clusters and the size of the emerging giant component. Link additions also allow cluster mergers and thus strongly affect cluster size distribution in a growing network. At the critical point, where the link addition rate reaches its critical value, there was a jump in average cluster size, and size distribution also changed abruptly. Link addition also affected the size of the giant component. The percolation was of infinite order, as the size of the emerging giant component depended on the link addition rate in a special way: all of the derivatives of the size of the giant component vanished as the addition rate converged to its critical value. In the opposite extreme of the parameter-set, when the duplication rate was extremely high, the network exhibited giant structural fluctuations in different realizations (Fig. **2**). This shows that mutations are vital for self-averaging and the emergence of robustness and statistical properties similar to those observed in real protein interaction networks.

The above observations are similar to our findings showing the importance of errors ('innovation') in the maintenance of cooperation in spatial games [34], and poses the very interesting opposite of Orgel's famous "error-catastrophe" model [57], implicating the existence of a 'perfection-catastrophe' on the other end. The required minimal rate of mutations and the appearance of critical behavior below this critical mutation rate can be interesting aspects of further studies concerning the evolution of protein structure, protein complexes and full-range protein interactions in cells.

In spite of the interesting advances above, we are far from the straightforward opportunity of functionally relevant, simultaneous percolation assessments of both protein structures and protein-protein interaction networks. Such a hierarchical percolation approach would allow the unbiased identification of key protein complexes and their hot-spots in the complex cellular architecture, and would also help us to understand the changes in information flow and importance in a large number of cellular states, including stress, cell cycle, cell differentiation, and various diseases.

# 6. INITIAL ATTEMPTS AND POSSIBLE WAYS TO MODEL PERTURBATION WAVES IN INTERACTOMES

Perturbation waves can efficiently model a number of key cellular processes, such as signal transduction, gene expression, as well as changes in metabolite concentrations. In the current review we restrict our summary to those, which can be interpreted as propagating conformational changes of proteins. The framework of perturbation waves can be applied to most processes above, and also allows the construction of efficient models to understand the propagation of noise, the fluctuations and diversity of individual cells governing phenotype variations, cellular movement, cell division and other physical rearrangements inside the cells and the threshold between smaller disturbances and larger damages [58,59].

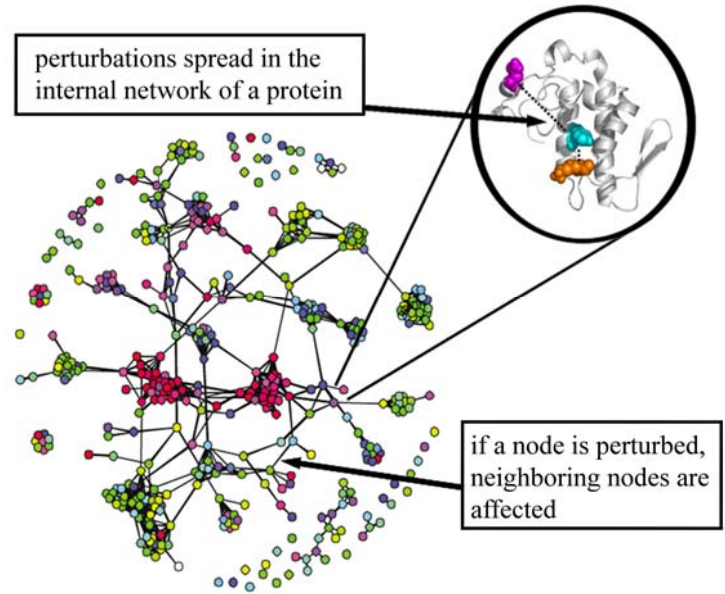

**Fig. (3).** Schematic representation of the hierarchical levels of perturbation-wave propagation in cellular networks. When modeling the propagation of perturbations we have both assess their spread inside the proteins (which can be well described by amino acids networks; for details, see text), and across the individual proteins, i.e. in the interactome of the cell.

Perturbation wave models have to consider the spread of perturbations inside and between individual proteins in the protein-protein interaction network (Fig. **3**). Prediction of distant effects after an initial perturbation applied to a component in a system of discrete elements is a rather difficult task. The spread of perturbation depends on several factors. First, the topology of the contact map (exemplified by both the amino-acid and protein-protein interaction networks) is decisive. As an example for the importance of network topology, protein interactomes having a small-world property dampen fluctuations and enhance synchronization [60]. Second, the coupling-characteristic is also an important factor. As key examples, coupling-delays and coupling strength both affect the stability of collective behavior [61]. Third, conditions required for a coherent action of network elements are grossly altered if the system is pulse-coupled (i.e. the interactions are pulse-like, where changes in the states of interacting elements are abrupt, large and return close to the original level within a short time). Many interactions between the elements of highly regulated biological systems can be described in terms of pulse-coupling [62]. Fourth, changes and evolution of contact-map topology are also important. In this latter case, snap-shots of evolutionary histories can also be averaged in time [63].

Coupled oscillator systems are commonly used to model the spread of perturbation in complex systems. In an effort to give account of both spatial and temporal variations of a complex system containing both attractions and repulsions moving oscillators have been studied in a bounded spatial domain with truncated elastic forces controlling their movements. In this model the interactions were restricted to an interaction range and were modified by the internal variables of the elements (e.g. by the phase differences between the oscillators). It has been shown that the emergence of clusters was largely affected by the collective phenomena of interaction ranges and the dispersion of time scales of changes in the internal variables [64]. The most relevant oscillator-related work to date assessing the propagation of perturbations considered a totally synchronized random network of phase-coupled oscillators and examined the effect of an external harmonic perturbation applied to one of them [65]. Other variables (spatial coordinates, delays, variable coupling strengths, etc.) were omitted in this study to give a first approach of the perturbation phenomenon. For small distances on a random network, the system was found to behave as a linear dissipative medium: the perturbation propagated at a constant speed, while its amplitude decreased exponentially with the distance. For larger distances, the response was saturated to an almost constant level, because the exponentially decaying signal had an exponentially growing number of available paths to reach the distant nodes. The results of this study can be extended to other interaction patterns and chaotic oscillators.

Collective behavior is not restricted to the simpler models of coupled oscillators, but also emerges in amino-acid networks of single protein molecules. Evidence from single-molecule experiments on a tetrameric enzyme ascertained the textbook model of cooperativity at a single-molecule level showing that ligand binding of the four subunits was not independent from each other: the majority of tetramers showed a one-step jump from no activity to the highest state of activity without observable intermediate states. Moreover, released inhibitors left occasional conformational traces behind causing heterogeneity among protein molecules [59]. So, it is straightforward to presume that complex communication-patterns emerge both inside individual proteins and in their oligomeric complexes. In an intriguing information-theoretic approach the communication characteristics of protein residues were investigated. In this discrete-time, discrete-state Markov model link affinities between residues were defined by the number of atom-atom contacts. Signal transmission probabilities were postulated to be proportional with link affinities. The hitting time was the average number of steps of the information residing at a residue to reach another in the amino-acid network of the protein. Communication processes were characterized with averages of hitting times. Functionally active residues were found to possess enhanced communication propensities. Additionally, a direct dependence between signal transduction events and equilibrium fluctuation dynamics has been detected [16]. As an additional example for the emergence of network-dependent 'hot-spots' in protein structure, a topology-based nonlinear network model of protein dynamics elicited that the spontaneous localization of vibrational energy is both wide-spread and site-dependent. In this model an original elastic network model (where the tension-derived energy is proportional with a second order term of atomic dislocations) was extended by adding a fourth order term of dislocations to the energy, which led to nonlinear forces. Nonlinear, high-frequency modes tended to be localized at the stiffest part of the network [66]. The localized vibrational energy stored in these regions may contribute to enzymatic activity. Thus local protein stiffness, central communication and enzymatic activity seem to be coupled. Vibrations of nonlinear origin may be concentrated at communicationally key residue-sets, and play an important role in energy storage and transfer during specific biological functions, like enzymatic processes or signal transduction. As a good additional example for the dynamical changes in perturbation-propagation, Ghosh and Vishveshwara [67] found four major communication paths in tRNA synthase, which were connected only, when the enzyme bound both substrates. This suggests the possibility of similar specialization of perturbation propagation pathways in case of special conditions requiring a specialized function instead of a general responsiveness. Such a scenario may occur at the cellular level during signaling.

Energy landscapes provide a coherent picture for the assessment of the energy-changes during conformational transitions. If energy minima are treated as nodes and possible transitions as links we get an 'energy-network', if different conformational states are the nodes linked by transitions than the resulting system is called as a 'conformational network'. Both can be useful for a complex representation of propagating perturbations, particularly after recent advances avoiding cases when different physical states have the same value of the measured observable [68]. A self-organized critical behavior often emerges in proteins, where relaxations are restricted. This is exemplified by the local accumulation of tensions and energy as shown in different models above, and by consecutive avalanches of propagating relaxations. Activation energies on energy landscapes are decreased by both water and molecular chaperones. The presence of water and chaperones 'softens' the network, make the propagation of perturbations smoother and their modeling easier [4,27,69]. Water and chaperones may play an important role in the 'fine-tuning' of enzyme activity and signal transduction.

At this point, we may conclude that exponential decays and smaller, rigid hot-spots with communication centrality are the only prevalent forms of interactions in coupled biological systems, and communication patterns are restricted to individual proteins or protein complexes. This is surely not the case. A number of examples, including the interactions of different receptors and actin filaments, show that conformational changes can propagate through extended lattices of protein molecules [58]. Duke et al. [70] offer an impressive theory motivated by statistical mechanics to explain these phenomena. They suppose a significant interaction between adjacent protein subunits, and also pose that two adjacent subunits having the same conformation have a lower combined energy, than the same adjacent subunits in different conformations. Treating ligand binding in energy terms, they obtain probabilities for different conformation states. Favorably, their model includes the canonical models of allostery as special cases. This obvious improvement does not come without a cost: the historical MWC (Monod, Wyman, Changeux) and KNF (Koshland, Néméthy, Filmer) models have two main parameters each. On the contrary, the general version of the Duke et al. model [70] contains five parameters, which makes it more complicated. Nevertheless, the general five-parameter model provides a comprehensive description of a variety of allosteric effects and is extendable to non-equilibrium states as well.

Dealing with real-world amino-acid and protein-protein interaction networks, there are two possibilities: either to use a complex simulation pattern aiming to encompass lots of different aspects (e.g. mapping the intracellular space onto a lattice, and fill it with moving proteins to capture the discrete and stochastic nature of interactions) [36], or to investigate single effects contributing to the understanding of the complex cellular behavior. An interesting example of the latter case is the work of Maslov and Ispolatov [5] computing responses to an abrupt, 2-fold local concentration change in the yeast protein-protein interaction network. The authors used the mass-action law to assess free protein concentrations in the yeast cell. The effects of a perturbation changing the abundance of a chosen protein were strongly localized: there was an exponential decay in the changes of free protein concentrations as moving away in the protein-protein interaction network from the perturbed node. Still, under specific favorable conditions concentration perturbations could selectively propagate over network distances up to four steps. Perturbations are certainly affected by the modular structure of both proteins and interactomes. Modules have been, in fact, defined by one approach as structures making the network more robust towards small perturbations [71]. The rationale behind this idea is, that once a perturbation reached a module, it 'gets lost' in the denser inter-modular link-structure, and has a smaller chance to affect adjacent modules. Both modular overlaps and the hierarchical organization of modules have large effects on the propagation of perturbations [72]. Thus changes of modular structure after various cellular events, such as stress [55] may greatly re-model perturbation pathways in cells.

Regretfully, a general model taking into account both the propagation of conformational changes inside the individual proteins as well as in the complex protein interactome is missing. Based on the above assumptions the following considerations have to be kept in mind when constructing such a model (Table **1**). Little is known of the propagation mechanisms of perturbations between proteins. Thus the currently existing conjectures about the possible ways of interaction do not allow the construction of a good model based on physical evidence. However, there are feasible phenomenological possibilities, where the key question is the proper form of coupling we should use. Although delays are often of great significance, biologically relevant pulse-like coupling highly complicates this problem, too. Models stemming from statistical mechanics (termed here as stochastic modeling) seem to deliver better results at this stage of research. Stochastic modeling offers another advantage: in such a model the protein structures and perturbation propagation in these amino acid networks are not studied. This is obviously a huge simplification, though it should be applied in all cases, where it is possible to achieve good agreement with experimental results without more rigorous investigations. If it is inevitable to analyze protein structures then models based on atomic connections are deemed to be the most effective. First, information-theoretical approaches are advantageous as they are easy to use and proper predictions for communication propensities can be expected. However, in the lack of detailed information about allosteric processes, the effect of large structural changes can not simply be incorporated, though it would be essential in case of most conformational spreads. Elastic models suffer the same problems: they are eligible for a proper description of equilibrium fluctuations, but they are unable to interpret protein folding. Neither do nonlinear models cope with this problem well: they are able to catch specific characteristics like energy localization, but the modified forces may lead to undesired effects and additional difficulties, and a proper description of allostery seems to be far away. Despite serious efforts, trajectories in the phase space of conformational states are unpredictable for a complete protein-folding event. In summary, it seems that further steps are needed to gain insight into perturbation propagation inside individual proteins allowing for a more precise modeling. Until then, we have to put up with coupled oscillator systems or more favorably with stochastic models of conformational spread.

**7. DISCRIMINATION BETWEEN SIGNALS AND NOISE**
Cellular networks function in an extremely noisy environment having both external and internal noise. Signals have to be learned by the amino acid networks and interactomes allowing an evolution of link-rearrangements, which provide an amplified 'highway' for signals, and filter noise. Signals are not only a learned property of networks, but their discriminatory network structures have to be special showing an inherent robustness against perturbations. As an often-studied example, the bacterial chemotactic pathway shows an optimally robust performance against perturbations while minimizing the cost of high protein abundance [73]. The yeast interactome shows another feature increasing robustness further at the systems level. Here dynamic modules with a higher flexibility for the condition-dependent regulation of cell behavior are segregated from static modules, which provide robustness to the cell against genetic perturbations or protein expression noise [40]. Network robustness is not 'free', not automatic. Millions of biological network variations studied by Ciliberti et al. [74] showed a skewed distribution, with a very small number of networks being vastly more robust than the rest. Very remarkably, these specifically robust networks were 'connected' and evolvable meaning that they can be easily transformed to each other by a small number of changes in network topologies. This property of biological networks gives a chance for the gradual evolution of signaling systems as skeletons of the underlying

Table 1. Gains and losses of possible perturbation wave models

| Model | Gain | Loss | Complications* |
|---|---|---|---|
| **Perturbation start** | | | |
| **Location of perturbation starting points** | Effects of different origins can be modeled | The same precision of data is needed in the whole network | * |
| **Perturbation waves with multiple starting points** | Possible model for interfering effects | Need for hardly available data on time dependence | ** |
| **Perturbation type** (single peak, peak-set, continuous, etc.) | Interaction among different perturbations | Need for detailed knowledge of all types | ** |
| **Shape of perturbation** (Gaussian, rectangle-like, sinusoid, etc.) | A greater complexity of perturbation events | Need for biological data of the shapes | ** |
| **Perturbation spread** | | | |
| **Directed and weighted coupling strength** | Strength of effects, directions of propagation | Growing computational costs | ** |
| **Non-zero coupling times, delays** | More precise models for time dependence | Need for precise biological data | ** |
| **No inner protein structure** | Easy to use | All structural effects | |
| *Topological models* | Interactome topology | Dynamical properties | * |
| *Stochastic model with statistical physical origins* | Experimental evidence on conformational spread | Need for correct parameter data (more or less available) | * |
| *Complex computer based solutions* | Discrete, stochastic behavior (interactions) | Complications swiftly increase for not so simple situations | * |
| **Simplified inner protein structure (oscillator models)** | Widely studied | One or few inner variables instead of structure | * |
| *Simple oscillator models* | Well known behavior | Biological resemblance | * |
| *Pulse-coupled oscillators* | One biological property regained | Need for parameter (e.g. delay) data (partially available) | ** |
| *Spatially moving oscillators* | Discrete interactions | Boundary effects; no studies concerning perturbation waves | ** |
| **Refined inner protein structure** | Reality (refined time dependence, exact propagation conduits) | Chances for describing allostery | *** |
| *Information theoretical* | Correct communication properties | Need for unavailable structure data | * |
| *Linear elastic models* | Correct predictions for equilibrium fluctuations | Need for unavailable structure data | ** |
| *Nonlinear elastic* | Description of energy localization | Need for unavailable structure data. Possible side-effects. | *** |
| **Effects of the medium** (water-induced fluctuations, cellular crowding-induced excluded volume, etc.) | Another important aspect regained | Complex interactions have to be considered | *** |

*The approximate level of difficulty has been marked by asterisks. Difficulties may come from increased computational complexity (longer run-times) as well as from the currently un-available data necessary for the particular method.

amino acid and protein networks [75], while preserving network robustness. The emergence of signal transduction pathway may use the following topological 'tricks':
- remodeling of link-density and link-weights resulting in the discrimination of roads and superhighways [75];
- linking and disjoining roads and superhighways allowing an efficient percolation of variable network regions [67];
- linking and disjoining hubs, thus constructing and destroying a rich-club [43];
- changes in the structure, overlaps and hierarchy of network modules [55,71,72].

We are at the very beginning of the understanding of the dynamical richness allowing the continuous emergence and suppression of the perturbation-channeling signaling topologies of our cells.

## 8. APPLICATIONS OF PERTURBATION WAVES IN DRUG DESIGN

Many drug-candidates, which have been designed to target a specific disease-related protein failed due to the intrinsic robustness of cellular networks against various perturbations [76]. Perturbation analysis of gene expression profiles as readily available systems level information in drug research has been successfully applied to identify drug-targets (primarily affected genes) with an approximate success rate of 70% to 80% [77]. The effects of perturbations on flux-balance analysis (metabolic control analysis) are also increasingly used for the identification of novel drug targets [78]. Signal transmission proteins have been increasingly identified as drug targets in the therapy of a large number of diseases [79,80]. However, we have only an initial understanding of changes in signal transduction pathways in stress, disease, or – actually – in the presence of a drug molecule causing resistance. The perturbation analysis methods outlined in this review will greatly help us to circumvent these formidable problems and to expand the currently rather small target-set in human proteins.

## 9. CONCLUSIONS AND PERSPECTIVES

In this review we have detailed the currently available applications of three powerful network-related methods, game theory, percolation theory and perturbation analysis on amino acid networks (i.e. single proteins), and protein-protein interaction networks (i.e. interactomes). In the following points we will briefly outline a few major findings and suggestions, which may be important for the progress in this field.

- **Protein games.** The identification of weak, long-range attractions [27] proved to be an important step for modeling 'protein-games' [29], i.e. the interdependence of mutually 'agreeable' conformational changes preparing two proteins to interact with each other. 'Fuzzyness' in terms of protein dynamics as well as structural disorder [35] together with 'learning' in terms of the communication and gradual changes described above may be crucial for the development of mutual cooperation [34] during the docking process.
- **Percolation.** Changes of percolation during the dynamical rearrangement of amino acid networks and protein-protein interaction networks, such as during protein folding or aging may offer a refined monitoring of the emergent properties of these complex systems. Different local levels of percolation may reveal different functional states of cellular modules/protein complexes, which may undergo profound changes in stress, during the cell cycle, cell differentiation and disease. Moreover, the simultaneous assessment of intra- and inter-protein percolation will give information on cellular dynamics at an unprecedented detail.
- **Propagation of perturbations.** When assessing the possible propagation of perturbations in cells, we have to take into account several layers of complexities (Table **1**). First, the starting conditions of perturbations have to be set, namely the number and location(s) of the starting point(s) of perturbation(s); the type of perturbation (e.g. single peak, peak-set, continuous, etc.) and finally, the shape of perturbation (e.g. Gaussian, rectangle-like, sinusoid, etc.). Second, coupling strength and coupling time (delays) have to be considered. Third, we have to decide, whether we will take into account the propagation of perturbations inside the proteins or not. If yes, whether we would like to use a grossly simplified or a detailed model. Fourth, we need to decide, if the effects of the medium (like water-induced fluctuations or cellular crowding) are included or not. If we will include all these complexities we will know practically anything on cellular dynamics. However, even half of the above complications go much beyond the current computational possibilities, and (more importantly) require much more than the currently available data. Despite of these difficulties modeling of perturbation propagation will be an extremely hot and extremely promising research are of the near future.

The above three powerful methods will be crucial in the design of novel drug targets and target-sets. This is especially true, if we want to avoid network remodeling-based resistance [76] and want to design drugs, which affect the targets of the grossly altered networks of 'sick cells'. We have to learn much more how cellular networks

(1) remodel their link-density and link-weights reshuffling network roads and superhighways [75];
(2) link and disjoin roads, superhighways and hubs allowing variable levels of percolation at different network regions [43,67]; and
(3) change the structure, overlaps and hierarchy of their modules [55,71,72].

We are at the very beginning of the understanding of the dynamics of cellular networks in stress, diseases and aging. The use of the above 'golden-triangle' of game theory, percolation theory and perturbation analysis may offer a winning strategy to get closer to the core of this problem.

**ACKNOWLEDGEMENTS**
Authors would like to thank members of the LINK-group (www.linkgroup.hu) for helpful suggestions. Work in the authors' laboratory was supported by research grants from the Hungarian National Science Foundation (OTKA-K69105) and from the EU (FP6-016003).